\documentclass{aa}
\usepackage{graphicx}
\usepackage{txfonts}
\usepackage{natbib}

\newcommand{\Msun}{{\rm M_{\odot}}}

\newcommand{\pc}{{\rm \, pc}}

\newcommand{\mpc}{{\rm \, Mpc}}
\newcommand{\kmps}{{\rm \, km \, s^{-1}}}

\newcommand{\K}{{\rm \, K}}

\newcommand{\Ha}{\rm H{\alpha}}
\newcommand{\Hb}{\rm H{\beta}}

\begin{document}
   \title{Starbursting Nuclear CO Disks of Early-type Spiral Galaxies}

   \subtitle{}

   \author{
J. Koda\inst{1,2,3,4} \and
T. Okuda\inst{2,5} \and
K. Nakanishi\inst{2} \and
K. Kohno\inst{5} \and
S. Ishizuki\inst{3} \and
N. Kuno\inst{2} \and
S. K. Okumura\inst{2}
          }

   \offprints{J. Koda}

   \institute{
JSPS Research Fellow; \email{koda@astro.caltech.edu} \and
Nobeyama Radio Observatory, Minamisaku, Nagano, 384-1305, Japan \and
National Astronomical Observatory, Mitaka, Tokyo, 181-8588, Japan \and
California Institute of Technology, MS 105-24, Pasadena, CA 91125, USA \and
Institute of Astronomy, University of Tokyo, Mitaka, Tokyo, 181-0016, Japan
             }

   \date{Received ; accepted 19/10/2004}

   \abstract{
We have initiated the first CO interferometer survey of early-type spiral
galaxies (S0-Sab). We observed five early-type spiral galaxies with HII
nuclei (indicating circumnuclear starburst activities). 
These observations indicate gas masses for the central kiloparsec
of $\sim 1-5\%$ of the dynamical masses. Such low gas mass fractions
suggest that large-scale gravitational instability in the gas is
unlikely to be the driving cause for the starburst activities.
We estimated Toomre $Q$ values and found that these
galaxies have $Q>1$ (mostly $>3$) within the central kiloparsec,
indicating that the gas disks are globally gravitationally stable.
From the brightness temperatures of the CO emission we estimated 
the area filling factor of the gas disks within the central kiloparsec
to be about 0.05. This small value indicates the existence of lumpy structure,
i.e. molecular clouds, in the globally-gravitationally stable disks.
The typical surface density of the molecular clouds is as high as 
$\sim3000\,\Msun\,\pc^{-2}$.
In the light of these new observations, we reconsider the nature of
the Toomre $Q$ criterion, and conclude that the Toomre $Q$ parameter
from CO observations indicates {\it neither} star formation
{\it nor} molecular cloud formation. This argument should be valid not
only for the circumnuclear disks but also for any region in galactic disks.
We tentatively explore an alternative model as an initiating mechanism
of star formation. Cloud-cloud collisions might account
for the active star formation.
   }

   \maketitle
%

\section{Introduction}



Early-type spiral galaxies (S0-Sab) generally have lower total gas masses
than late-type spiral galaxies (Sb-Scd), and their global star formation
rates (SFR) are generally less than those
of late-type spirals \citep{you91,rob94}.
Circumnuclear starbursts (HII nuclei) are also less frequent in
early-type spiral galaxies than in late-type spiral galaxies. However,
when early-type spiral galaxies {\it do} host circumnuclear starbursts,
their extinction-corrected $\Ha$ luminosities (measuring SF activity)
are much higher on average than those of late-type spiral galaxies
\citep{ho97a, ken98}.
Similar trends are also seen in $\rm 10 \mu m$ surveys of nearby
galaxies \citep{dev84, giu94}.

High resolution CO observations are critical for understanding
the active nuclear SF in early-type spiral galaxies.
Although several high resolution surveys of molecular gas in
galactic centers have recently been completed \citep{sak99a, sof03a,
tam03, gar03}, they include very few early-type spiral galaxies,
especially with HII nuclei. This paper presents the first
results from a CO survey of early-type spiral galaxies with HII nuclei
using the Nobeyama Millimeter Array (NMA), specifically addressing
the region of the nuclear starburst activity.



\section{Samples and CO Observations}

%
%
%
%
%

Our sample of early-type spiral galaxies is selected from the
optical-spectroscopic survey of \citet{ho97a} with the criteria:
(a) morphological type of S0-Sab,
(b) hosting an HII-nucleus
[Note that HII nucleus here means a circumnuclear region with SF
activity but without AGN],
(c) not edge-on [$i <~70^{\circ}$],
(d) distance $< 25\mpc$, and
(e) no obvious interaction.
From the complete sample of 11 galaxies satisfying these conditions,
we have observed 5 galaxies for the first time and found one already
observed and reported (Table \ref{tab:sample}). [The other five
galaxies are NGC 3073, 4245, 4424, 4448, and 4694.]

Table \ref{tab:sample} also includes $\Ha$ equivalent widths, and star
formation rates (SFR) within the central 1 kpc.
The SFR is calculated as in \citet{sak99b} from the $\Ha$-photometry
with $4\arcsec \times 2\arcsec$ aperture by \citet{ho97b}.
We made the extinction correction using the $\Ha/\Hb$ line intensity ratio,
calculated the $\Ha$ luminosity from the central kiloparsec region
assuming constant line intensity
in the area, and used the $\Ha$-to-SFR conversion formula \citep{ken98}.
This procedure usually gives SFRs consistent with those based
on infrared observations \citep{sak99b}. There are, however,
some caveats for these optical SFRs. They are likely to be
lower limits, because some $\Ha$-emission from embedded SF in dust
may have been missed, and because absorption in the background stellar
continuum may lead to underestimation of the $\Ha$-luminosity.
The HII nuclei do not host AGNs by definition, and therefore
there is no contamination of AGNs in the estimation of SFR.

\begin{center}
------ Table 1 ------
\end{center}

%
%

$^{12}$CO($J=1-0$)-line observations  were obtained using NMA in November,
2002 through May, 2003. Each object was observed for two 10-hour
tracks in the C- and D-array configurations of NMA.
The complex gain was calibrated every 20 minutes using
nearby quasars. The fluxes of the quasars were calculated against URANUS;
uncertainties in the overall flux scales are $\sim 15\%$.
The visibility data were calibrated using the UVPROC-II
software package \citep{tsu97} and mapped with CLEAN in
the NRAO/AIPS package. The resultant 3-D cubes
have resolutions of $4-5\arcsec$ (natural weighting)
and rms noise of $\sim 17{\rm \,mJy/beam}$ in a $10.4 \kmps$ channel.
(Full channel maps and kinematic maps will be presented in our
full survey paper). These interferometric maps include $\gtrsim 85\%$
of the total single dish CO line fluxes (Okuda, in preparation)
for the central $15\arcsec$ region in each galaxy.

\section{Molecular Gas in Starbursting Nuclear Disks}

\begin{figure}
   \centering
   \includegraphics[width=\textwidth]{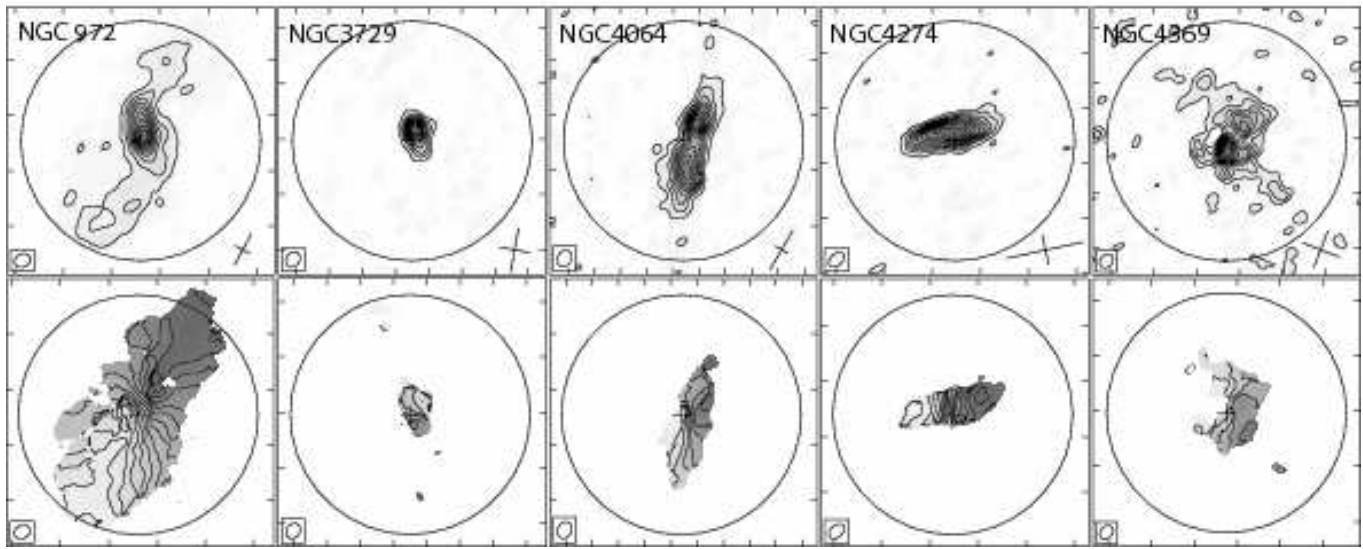}
      \caption{
{\it Upper panels:} CO integrated intensity maps. Contours
represent (0.1, 0.2, 0.3, 0.4, ..., 0.9, 1.0) $\times$ peak
intensity. The peak intensities are 58.2, 27.6, 14.9, 26.5,
8.57 $\rm \, Jy \, beam^{-1} km \, s^{-1}$ from left to right.
Circles and crosses indicate the fields of view ($65\arcsec$)
and their centers in the observations, respectively.
Crosses at the lower right corners
represent 1 kpc along major and minor axes. No primary beam
correction has been applied to these maps.
{\it Lower panels:} velocity field maps. These maps are made
using $>3\sigma$ emissions in each $10.4\kmps$ channel.
Contour intervals are $10\kmps$ for NGC 4369 and $20\kmps$ for
the others. Dark sides are receding.
              }
         \label{fig:face}
   \end{figure}

We detect significant CO emission in the circumnuclear regions
of all 5 early-type spiral galaxies. Figure \ref{fig:face} shows
CO integrated intensity maps and velocity fields [NGC 3593, the
sixth galaxy in our sample, is reported in \citet{sak99a}].
In all 6 galaxies, the CO emission is strongly peaked in the central
region at $R<500\rm pc$. Linear scales of 1 kpc along major and minor
axes are indicated by the crosses in Fig. \ref{fig:face}.
In Table \ref{tab:sample} we include estimates for the dynamical
mass ($M_{\rm dyn}$), gas mass ($M_{\rm gas}$), and average gas surface
density ($\Sigma_{\rm gas}$) within a radius of 500 pc from
the dynamical center of each galaxy. The gas mass is estimated using
CO fluxes within the central 1 kpc aperture on the galactic disks, and
a Galactic CO-to-H$_2$ conversion factor of $X_{\rm CO} = 3.0 \times
10^{20} \,\rm cm^{-2} (K \, km\, s^{-1})^{-1}$ and a factor of 1.36 to
account for He and the other elements. The dynamical mass is calculated
using the formula $M_{\rm dyn}=R V^2/G$. The velocity at $R=500\pc$ is
estimated from the velocity difference between $R =\pm 500\pc$
along the major axis in position-velocity diagrams.

\begin{center}
------ Figure 1 ------
\end{center}


The circumnuclear regions in these galaxies are regarded as starburst
regions by two definitions.
First, the gas consumption timescales, $M_{\rm gas}/\rm SFR\sim 2 \times
10^8\,\rm yr$ on average, are as short as those of starburst galaxies
\citep{ken98}. Second, the current SFRs $\sim 0.7\,\Msun\,\rm yr^{-1}$
are larger than past-averaged SFRs, i. e. $M_{\rm star}/10^{10} \,\rm yr \sim
0.3\,\Msun\,\rm yr^{-1}$ (assuming $M_{\rm star}\sim M_{\rm dyn}$).
In addition, the $\Ha$ equivalent widths, $\sim 36 \,\AA$, are
much larger than typical values $\sim 0-10\,\rm \AA$ for early-type spiral
galaxies \citep{ken98}, indicating that their current SFRs are higher than
most other galaxies.


Despite the starbursts, the average gas surface densities at $R<500\pc$
are lower in the early-type spiral galaxies, $\sim 150\,\Msun\,\pc^{-2}$,
than in late-type spirals with and without HII nuclei
\citep[mostly $>200\,\Msun\,\pc^{-2}$, see ][]{sak99b}.
Even the non-starbursting interacting galaxy NGC 5195
has a much higher surface density of $500\,\Msun\,\pc^{-2}$ at $R<500\pc$
\citep{koh02}. {\it Thus, the large-scale gas surface density is 
not the only determinant of nuclear SF.}
These global surface densities are very close to, or higher than,
the average surface density of molecular clouds, $170\,\Msun\,\pc^{-2}$,
in the Galactic disk \citep{sol87}. They, however, are not
a sufficient condition for initiating SF.

%
%

\begin{figure}
   \centering
   \includegraphics[width=8cm]{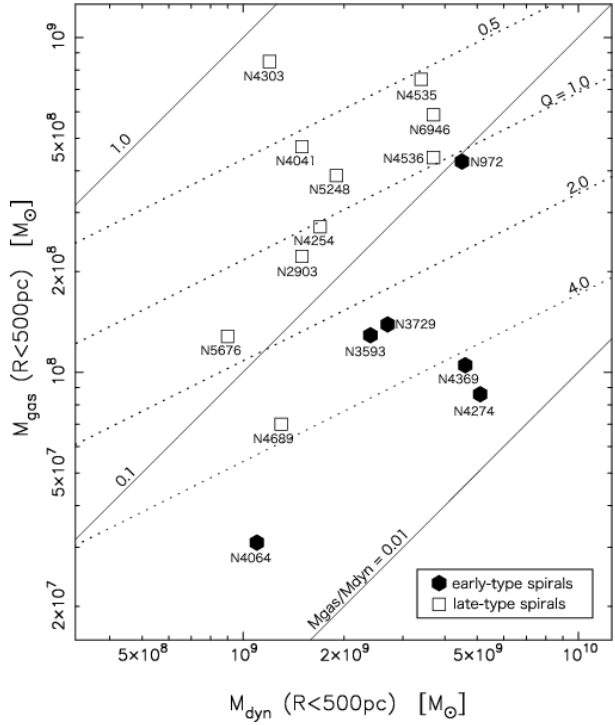}
      \caption{
Molecular gas mass vs. dynamical mass within 500 pc of
the nucleus. A Galactic CO-to-H$_2$ conversion factor of
$X_{\rm CO} = 3.0 \times 10^{20} \,\rm cm^{-2} (K \, km\, s^{-1})^{-1}$
is adopted.
Solid lines indicate constant gas-to-dynamical mass ratios.
Dotted lines indicate constant Toomre $Q$-values
assuming rigid rotation curve and velocity dispersions of
$\sigma = 10\kmps$. $Q>1$ indicates gravitationally stable,
while $Q<1$ indicates gravitationally unstable. The $Q$-values
become larger if we assume larger $\sigma$ (i.e. $Q \propto\sigma$)
and smaller $X_{\rm CO}$, which are more likely in the central region.
              }
         \label{fig:comp}
   \end{figure}

Figure \ref{fig:comp} shows the molecular gas masses plotted
against dynamical masses at $R< 500 \pc$.
Late-type spiral galaxies (Sbc-Scd) with HII nuclei
\citep{sak99b,sof03a} are also plotted.
The solid lines indicate constant gas-to-dynamical
mass ratios. The early-type spiral galaxies are distributed
between $M_{\rm gas}/M_{\rm dyn}=0.01$ and 0.1 (mostly in the range
0.02-0.05).
Evidence has accumulated for a lower conversion factor $X_{\rm CO}$ in
metal rich regions such as galactic centers \citep{wil95, ari96}.
If we adopt an $X_{\rm CO}$ that is three times as small as the
assumed one,
$M_{\rm gas}/M_{\rm dyn}$ becomes three times as low.
Given these low gas mass fractions,
{\it it is unlikely that the circumnuclear gas in these early-type
spiral galaxies has undergone large-scale gravitational collapse
and that this initiated the nuclear starbursts}
\citep[see ][]{jog84, wad92}.
Our conclusion that this sample of early-type spiral galaxies has low
$M_{\rm gas}/M_{\rm dyn}$ is very different from the result by
\citet{sak99b} who found that the later-type spiral galaxies 
with HII nuclei mostly have $M_{\rm gas}/M_{\rm dyn}>0.1$
(see Fig. \ref{fig:comp}).

In order to quantify the gravitational stability of the gas disks,
we estimate the Toomre $Q$ parameter averaged within central 1 kpc.
We adopt the formula suggested in \citet{sak99b} with a small
modification. $Q$ is described as
\begin{equation}
Q = \alpha \frac{(M_{\sigma} M_{\rm dyn})^{1/2}}{M_{\rm gas}},
\end{equation}
where $M_{\sigma} \equiv R \sigma^2/G$ and $\sigma$ is the velocity
dispersion. We assumed a rotation curve of $V \propto R^{\beta}$,
i.e. $\beta=1$ for rigid rotation and $\beta=0$ for flat rotation,
and then, $\alpha = \sqrt{2 (\beta + 1)}$.
We notice that $\alpha$ (and thus $Q$) is insensitive to the
rotation curve, and between 1.4 (flat rotation)
and 2 (rigid rotation).
The dotted lines in Fig. \ref{fig:comp} show constant $Q$ values
in assuming
$\sigma =10 \kmps$ and $\alpha=2$ \citep[similar to ][]{sak99b}.
As already suggested by the small values of $M_{\rm gas}/M_{\rm dyn}$,
all the early-type spiral galaxies have $Q>1$ (mostly $Q\gtrsim 3$),
indicating that the circumnuclear gas disks
are globally gravitationally-stable despite their clear starburst
activity.
We note that $\sigma$ is likely to be larger than our assumed value
of $10\kmps$ \citep{ken93}. If we adopt $\sigma=30\kmps$ and three
times as small $X_{\rm CO}$ \citep{ari96}, the $Q$s become nine
times as large, resulting in $Q\gtrsim 27$ (i.e. much more stable).


\begin{center}
------ Figure 2 ------
\end{center}

%

The peak brightness temperature $T_{\rm b}$ in the channel maps is $1-2\K$.
Assuming that the CO gas is optically thick as observed in molecular
clouds in the Galaxy \citep{ss87} and that the gas excitation
temperature $T_{\rm ex}$ is similar to the dust temperature of 30$\K$
in the Galactic center \citep{gau84},
the molecular gas area filling factor is $f\sim 0.05$
(=$T_{\rm b}/T_{\rm ex}-T_{\rm CMB}$), where the cosmic microwave
background temperature is $T_{\rm CMB}\sim 3\K$.
The filling factor could be lower if the dust temperature is higher.
{\it Therefore, the molecular gas is not uniformly
distributed but already confined in molecular clouds, although
the gas disk is globally gravitationally-stable.}
The typical gas surface density of the clouds becomes as high as
$\sim 3000\,\Msun\,\pc^{-2}$ ($=\Sigma_{\rm gas}^{500}/f$). This is
much larger than the average value, $170\,\Msun\,\pc^{-2}$,
for molecular clouds in the Galactic disk and close to the value,
$2500\,\Msun\,\pc^{-2}$, for the ones in the Galactic center
\citep[][ calculated as in Solomon et al. 1987]{oka01}.
The presence of molecular clouds in a stable disk is understood,
if the disk was gravitationally unstable in the past and heated
up after molecular cloud formation. Gravitational interactions
among clouds may suffice to increase cloud-cloud velocity dispersions
and heat up the global disk \citep{jog88}.

\section{Discussion}

Although we initially expected to detect significantly large gas masses
so that gravitational instability within the stellar potential might
account for the starbursting, our survey reveals that the early-type
spiral galaxies with circumnuclear starbursts have extremely low
$M_{\rm gas}/M_{\rm dyn}$ in the central kpc.
Thus large-scale gravitational instability is not a likely cause of
the starbursts.
Although Toomre $Q$-type gravitational instability is often used for
interpreting SF in galactic disks, its relevance for the circumnuclear
SF of early-type spiral galaxies is unsupported
\citep[see also, ][]{tho95,won02,boi03}.
These results may suggest reconsidering the nature of the Toomre $Q$.

If CO is detected, molecular clouds must already exist
unless the mass function of molecular clouds is extremely different
from those in the Galactic disk \citep{sco87,sol87} and in the
Galactic center \citep{oka01}; we have shown the existence of
dense molecular clouds in our sample.
Hence $Q$ from CO observations does not indicate molecular cloud formation.
In $Q$-value calculations, a velocity dispersion among molecular clouds
$\sim 10\kmps$ (an effective sound speed $c_{\rm s}$ for $\sim 10^4\K$) is
usually adopted, rather than the molecular gas temperature ($\sim 10\K$,
or $c_{\rm s} \sim 0.3\kmps$),
although the gas temperature should be more relevant for the
collapse of protostellar cores.
Thus, $Q$ cannot indicate SF directly. Instead, $Q$ obtained in earlier
studies might indicate the formation of giant molecular cloud
association (GMA), i.e. a group of molecular clouds.
Therefore, GMA formation has been implicitly assumed to be
necessary and sufficient conditions for SF in earlier Toomre $Q$
arguments.
Considering this assumption, we are not confident that the Toomre
$Q$-type gravitational instability is relevant for SF.
[Note, however, that a low $Q$-value means a relatively
high gas surface density. Thus, active SF is more likely to be
found in low $Q$-value regions if it is initiated by some other
mechanisms than the Toomre $Q$-type instability.]
The above argument should be valid not only for circumnuclear
disks but also for any region in galactic disks.
Perhaps the internal structure of molecular clouds, e.g.
lumpiness and/or turbulence, is more important for star formation
than large-scale gravitational instability.


One would think that massive bulges in early-type spiral
galaxies could be a key to understanding the galaxy-type dependence
of circumnuclear SF activity. In fact, \citet{dej96} found that
the bulge effective radius is 100-300 pc, while the disk scalelength
is 1-5 kpc. Hence bulge potentials could significantly affect gas
dynamics within the central kpc, and circumnuclear rotation curves
would be flatter in early-type spiral galaxies than in late-type
spirals. A flat rotation curve causes a high shear velocity due
to differential rotation. For example, assuming an angular velocity
$\Omega=1\kmps\,\pc^{-1}$, the extension of a molecular cloud $a=20\pc$
and the amplitude of the epicyclic motion $l=10\pc$ ($\approx \sigma/\kappa$;
random velocity $\sigma\sim10\kmps$ and epicyclic frequency
$\kappa\sim 1.4\Omega$), the shear velocity, $v=(\Omega-dV/dR)(a+l)$,
becomes as high as $30\kmps$ in a flat rotation curve, but $0\kmps$
in a rigid rotation curve. Thus the shear velocity dominates the random
velocity dispersion in a flat rotation curve. These facts, in addition to
the existence of high density molecular clouds in our sample,
leads us to speculate on cloud-cloud collisions induced by galactic
shear as a cause of starbursting in early-type spiral galaxies.
Obviously, cloud collision will enhance lumpiness and star
formation in molecular clouds, although the molecular clouds may not
contract globally in the strong shear regions.

In the Galaxy there is considerable evidence that the existence
of molecular clouds is necessary but not sufficient for SF \citep{moo88,
sco89}, and that cloud-cloud collisions might initiate active SF
in many instances \citep{lor76,sco86,ode92,has94}.
For a thin disk (2-D), the timescale for cloud collision is
$t_{\rm coll} = 1/n a v$, where $n=\Sigma_{\rm gas}/M_{\rm C}$
is the typical surface number density of molecular clouds,
$M_{\rm C}$ and $a$ are typical cloud mass and diameter,
and $v$ is the velocity of a test cloud relative to field clouds.
We here adopt $M_{\rm C}=10^6 \Msun$, $a=20\pc$,
$\Sigma_{\rm gas}=150\, \Msun \, \pc^{-2}$.
These parameters indicate that the surface density of individual
molecular clouds is $3000\,\Msun\,\pc^{-2}$, which is close
to the observed value. If we assume a shear velocity of
$v=30\kmps$, the collisional timescale becomes $t_{\rm coll} \sim
10^7$ yr. Given this timescale, the observed SFR of
$\sim 1 \, \Msun\,\rm yr^{-1}$ is achieved even if the SF efficiency
per collision is as low as 10 \%. This is consistent with
the low SF efficiencies in molecular clouds observed in the Galaxy.
Therefore, the cloud collision model induced by galactic shear
can reasonably explain the observed SF properties in our sample.
Note that the assumed cloud has a tidal diameter of
$\sim 20-150\pc$ at $R=100-500\pc$, and thus survives against
stellar gravitational potentials. This tentative idea will be
revisited in our full survey paper.


The results from our first CO survey of early-type spiral galaxies
require a mechanism for triggering active SF in low gas density
regions. The mechanism cannot be the Toomre $Q$-type gravitational
instability, and we have discussed a tentative model based on
cloud-cloud collisions. We are extending the CO survey with
the Nobeyama Millimeter Array and Nobeyama 45m telescope. We will
observe all early-type spiral galaxies with central starburst
in \citet{ho97a}. In addition to the CO survey, higher resolution
observations of the distribution of SF regions within the central
kpc at infrared wavelengths, and observations of
physical conditions in the gas at some other molecular lines,
e.g. rare CO isotopes and HCN (denser gas tracer, i.e. proximate
sites of SF), will be complementary to fully understand the
circumnuclear SF.

\begin{acknowledgements}
We thank the NMA staff for helping observations.
J. K. thanks Nick Scoville for deep discussions, Zara Scoville for
suggestions on the manuscript, and Sachiko Onodera and
Antonio Garcia-Barreto for comments on the manuscript.
We also thank an anonymous referee for useful comments.
J. K. and K. N. were financially supported by the Japan Society
for the Promotion of Science (JSPS) for Young Scientists.
This work was partly supported by a Grant-in-Aid for Scientific
Research (no. 15037212).

\end{acknowledgements}

\clearpage


   \begin{table}
      \caption[]{Sample Spiral Galaxies with HII Nuclei}
         \label{tab:sample}
     $$
         \begin{array}{llcccccccccc}
            \hline \hline
 & & \multicolumn{1}{c}{d} & \multicolumn{1}{c}{M_{B_T}^0} & \multicolumn{1}{c}{D_{25}^0} & \multicolumn{1}{c}{\rm EW(\Ha)} & \multicolumn{1}{c}{\rm SFR^{500}} & \multicolumn{1}{c}{M_{\rm gas}^{500}} &  \multicolumn{1}{c}{M_{\rm dyn}^{500}} &  \multicolumn{1}{c}{f_{\rm gas}^{500}}&  \multicolumn{1}{c}{\Sigma_{\rm gas}^{500}}&  \multicolumn{1}{c}{}\\
\multicolumn{1}{c}{\mbox{Name}} & \multicolumn{1}{c}{\mbox{Morph.}}& \multicolumn{1}{c}{\mbox{(Mpc)}} & \multicolumn{1}{c}{\mbox{(mag)}} & \multicolumn{1}{c}{\mbox{(kpc)}} & \multicolumn{1}{c}{\mbox{(\AA)}}  & \multicolumn{1}{c}{(\Msun \, {\rm yr}^{-1})} & 
\multicolumn{1}{c}{(10^7 \Msun)} & \multicolumn{1}{c}{(10^9 \Msun)} &  \multicolumn{1}{c}{(\%)} &  \multicolumn{1}{c}{(\Msun \, \pc^{-2})} & \multicolumn{1}{c}{\mbox{Reference}} \\
\multicolumn{1}{c}{(1)} & \multicolumn{1}{c}{(2)} & \multicolumn{1}{c}{(3)} & \multicolumn{1}{c}{(4)} & \multicolumn{1}{c}{(5)} & \multicolumn{1}{c}{(6)} & \multicolumn{1}{c}{(7)} & \multicolumn{1}{c}{(8)} & \multicolumn{1}{c}{(9)} & \multicolumn{1}{c}{(10)} & \multicolumn{1}{c}{(11)} &  \multicolumn{1}{c}{(12)} \\
            \hline
\mbox{NGC 972 } &\mbox{SAab      }&  21.4&  -20.17&  22.10 &  34.22 & 1.58 &  42.6 &  4.5 & 9 & 542 & \mbox{This work}\\
\mbox{NGC 3593} &\mbox{SA(s)0/a: }&   5.5&  -17.25&   7.84 &  13.13 & 0.28 &  12.9 &  2.4 & 5 & 164 & \mbox{KS99}     \\
\mbox{NGC 3729} &\mbox{SB(r)apec }&    17&  -19.24&  13.95 &  76.74 & 0.99 &  13.9 &  2.7 & 5 & 177 & \mbox{This work}\\
\mbox{NGC 4064} &\mbox{SB(s)a:pec}&  16.9&  -19.39&  21.97 &  55.73 & 0.43 &   3.1 &  1.1 & 3 &  39 & \mbox{This work}\\
\mbox{NGC 4274} &\mbox{(R)SB(r)ab}&   9.7&  -19.35&  19.53 &  4.65  & -    &   8.6 &  5.1 & 2 & 109 & \mbox{This work}\\
\mbox{NGC 4369} &\mbox{(R)SA(rs)a}&  21.6&  -19.40&  13.13 &  29.76 & 0.39 &  10.5 &  4.6 & 2 & 134 & \mbox{This work}\\
            \hline
         \end{array}
     $$
\begin{list}{}{}
\item[Note:]
(1) Galaxy name; 
(2) Morphological type;
(3) Distance based on Virgo infall model of \citet{tul84};
(4) Total absolute $B$-band magnitude corrected for Galactic and internal reddening and redshift from Ho et al. (1997b);
(5) Face-on isophotal diameter at $\mu_{\rm B}=25 {\,\rm mag\,arcsec^2}$ in the $B$-band, corrected for extinction and inclination from RC3 (de Vaucouleurs et al. 1991);
(6) $\Ha$ equivalent width;
(7) Star formation rate within 500 pc of the nucleus;
(8) Gas mass within 500 pc of the nucleus, from CO interferometer observations assuming a conversion factor of $X_{\rm CO} = 3.0 \times 10^{20}{\rm \,cm^{-2}\,(K\,km\,s^{-1})^{-1}}$;
(9) Dynamical mass within 500 pc of the nucleus;
(10) Gas-to-dynamical mass ratio within 500 pc from the center;
(11) Surface gas density averaged within 500 pc from the center;
(12) Reference for CO interferometer observations. KS99: Sakamoto et al. (1999b).
\end{list}
\end{table}

\clearpage

\clearpage

\begin{thebibliography}{}
\bibitem[Arimoto et al.(1996)]{ari96} Arimoto, N., Sofue, Y. \& Tsujimoto, T. 1996, \pasj, 48, 275
\bibitem[Boissier et al.(2003)]{boi03} Boissier, S., Prantzos, N., Boselli, A. \& Gavazzi, G. 2003, \mnras, 346, 1215
\bibitem[de Jong(1996)]{dej96} de Jong, R. S. 1996, \aa, 313, 45
\bibitem[Devereux(1984)]{dev84} Devereux, N. 1984, \apj, 323, 91 
\bibitem[Garcia-Burillo et al.(2003)]{gar03} Garcia-Burillo, S., Combes, F., Hunt, L. K. et al. 2003, \aap, 407, 485
\bibitem[Gautier et al.(1984)]{gau84} Gautier, T. N., Hauser, M. G., Beichman, C. A., Low, F. J. et al. 1984, \apjl, 278, 57
\bibitem[Giuricin et al.(1994)]{giu94} Giuricin, G., Tamburini, L., Mardirossian, F., Mezzetti, M., \& Monaco, P. 1994, \apj, 427, 202
\bibitem[Hasegawa et al.(1994)]{has94} Hasegawa, T., Sato, F., Whiteoak, J. B. \& Miyawaki, R. 1994, \apjl, 429, 77
\bibitem[Helfer et al.(2003)]{tam03} Helfer, H. T., Thornley, M. D., Regan, M. W. et al. 2003, \apjs, 145, 259
\bibitem[Ho, Filippenko \& Sargent(1997a)]{ho97a} Ho, L. C., Filippenko, A. V., \& Sargent, W. L. W. 1997a, \apj, 487, 579
\bibitem[Ho, Filippenko \& Sargent(1997b)]{ho97b} Ho, L. C., Filippenko, A. V., \& Sargent, W. L. W. 1997b, \apjs, 112, 315
\bibitem[Jog \& Ostriker(1988)]{jog88} Jog, C. J. \& Ostriker, J. P. 1988 \apj, 328, 404
\bibitem[Jog \& Solomon(1984)]{jog84} Jog, C. J. \& Solomon, P. M. 1984 \apj, 276, 114
\bibitem[Kenney et al.(1993)]{ken93} Kenney, J. D. P., Carlstrom, J. E. \& Young, J. S. 1993, \apj, 418, 687
\bibitem[Kennicutt(1989)]{ken89} Kennicutt, R. C. 1989, \apj, 344, 685
\bibitem[Kennicutt(1998)]{ken98} Kennicutt, R. C. 1998, \araa, 36, 189
\bibitem[Kohno et al.(2002)]{koh02} Kohno, K., Tosaki, T., Matsushita, S., Vila-Vilao, B., Shibatsuka, T., \& Kawabe, R. 2002, \pasj, 54, 541
\bibitem[Loren(1976)]{lor76} Loren, R. B. 1976, \apj, 209, 446
\bibitem[Magorrian et al.(1998)]{mag98} Magorrian, J., Tremaine, S., Richstone, D., Bender, R., et al. \aj, 115, 2285
\bibitem[Mooney \& Solomon(1988)]{moo88} Mooney, T. J. \& Solomon, P. M. 1988, \apjl, 334, 51
\bibitem[Odenwald et al.(1992)]{ode92} Odenwald, S., Fischer, J., Lockman, F. J. \& Stemwedel, S. 1992, \apj, 397, 174
\bibitem[Oka et al.(2001)]{oka01} Oka, T., Hasegawa, T., Sato, F., Tsuboi, M., Miyazaki, A. \& Sugimoto, M. 2001, \apj, 562, 348
\bibitem[Roberts \& Haynes(1994)]{rob94} Roberts, M. S. \& Haynes, M. P. 1994, \araa, 32, 115
\bibitem[Sakamoto et al.(1999a)]{sak99a} Sakamoto, K., Okumura, S. K., Ishizuki, S., \& Scoville, N. Z.  1999a, \apjs, 124, 403
\bibitem[Sakamoto et al.(1999b)]{sak99b} Sakamoto, K., Okumura, S. K., Ishizuki, S., \& Scoville, N. Z.  1999b, \apj,  525, 691
\bibitem[Scoville \& Good(1989)]{sco89} Scoville, N. Z. \& Good, J. C. 1989, \apj, 339, 149
\bibitem[Scoville \& Sanders(1987)]{ss87} Scoville, N. Z. \& Sanders, D. B. 1987, Interstellar Process, ed. D. Hollenbach and H. Thronson (Durdrecht: Reidel).
\bibitem[Scoville et al.(1987)]{sco87} Scoville, N. Z., Yun, M. S., Clemens, D. P., Sanders, D. B. \& Waller, W. H. 1987, \apjs, 63, 821
\bibitem[Scoville, Sanders \& Clemens(1986)]{sco86} Scoville, N. Z., Sanders, D. B. \& Clemens, D. P. 1986, \apjl, 310, 77
\bibitem[Sofue et al.(2003)]{sof03a} Sofue, Y., Koda, J., Nakanishi, H., Onodera, S., Kohno, K. \& Tomita, A. 2003, \pasj, 55, 17
\bibitem[Solomon et al.(1987)]{sol87} Solomon, P. M., Rivolo, A. R., Barrett, J. \& Yahil, A. 1987, \apj, 319, 730
\bibitem[Thornley \& Wilson(1995)]{tho95} Thornley, M. D. \& Wilson, C. D. 1995, \apj, 447, 616
\bibitem[Toomre(1964)]{too64} Toomre, A. 1964, \apj, 139, 1217
\bibitem[Tsutsumi et al.(1997)]{tsu97} Tsutsumi, T., Morita, K. -I., \& Umeyama, S. 1997, in ASP Conf. Ser. 125, Astronomical Data Analysis Software and Systems VI, ed. G. Hunt \& H. E. Payne (San Francisco; ASP), 50
\bibitem[Tully \& Shaya(1984)]{tul84} Tully, R. B., \& Shaya, E. J. 1984, \apj, 281, 31
\bibitem[Wada(1992)]{wad92} Wada, K. \& Habe, A. 1992, \mnras, 258, 82
\bibitem[Wilson(1995)]{wil95} Wilson, C. D. 1995, \apjl, 448, 97
\bibitem[Wong \& Blitz(2002)]{won02} Wong, T. \& Blitz, L. 2002, \apj, 569, 157
\bibitem[Young \& Scoville(1991)]{you91} Young, J. S. \& Scoville, N. Z. 1991, \araa, 29, 581
\end{thebibliography}
\end{document}